\documentclass[twocolumn, prl,aps,showpacs,superscriptaddress]{revtex4-1}
\usepackage{graphicx}
\usepackage{amsmath}

\begin{document}

\title{Evaporative Cooling of Antiprotons to Cryogenic Temperatures}
\author{G.B. Andresen}
\affiliation{Department of Physics and Astronomy, Aarhus University, DK-8000 Aarhus C, Denmark}
\author{M.D. Ashkezari}
\affiliation{Department of Physics, Simon Fraser University, Burnaby BC, V5A 1S6, Canada}
\author{M. Baquero-Ruiz}
\affiliation{Department of Physics, University of California at Berkeley, Berkeley, CA 94720-7300, USA}
\author{W. Bertsche}
\affiliation{Department of Physics, Swansea University, Swansea SA2 8PP, United Kingdom}
\author{P.D. Bowe}
\affiliation{Department of Physics and Astronomy, Aarhus University, DK-8000 Aarhus C, Denmark}
\author{E. Butler}
\affiliation{Department of Physics, Swansea University, Swansea SA2 8PP, United Kingdom}
\author{C.L. Cesar}
\affiliation{Instituto de F\'{i}sica, Universidade Federal do Rio de Janeiro, Rio de Janeiro 21941-972, Brazil}
\author{S. Chapman}
\affiliation{Department of Physics, University of California at Berkeley, Berkeley, CA 94720-7300, USA}
\author{M. Charlton}
\affiliation{Department of Physics, Swansea University, Swansea SA2 8PP, United Kingdom}
\author{J. Fajans}
\affiliation{Department of Physics, University of California at Berkeley, Berkeley, CA 94720-7300, USA}
\author{T. Friesen}
\affiliation{Department of Physics and Astronomy, University of Calgary, Calgary AB, T2N 1N4, Canada}
\author{M.C. Fujiwara}
\affiliation{TRIUMF, 4004 Wesbrook Mall, Vancouver BC, V6T 2A3, Canada}
\author{D.R. Gill}
\affiliation{TRIUMF, 4004 Wesbrook Mall, Vancouver BC, V6T 2A3, Canada}
\author{J.S. Hangst}
\affiliation{Department of Physics and Astronomy, Aarhus University, DK-8000 Aarhus C, Denmark}
\author{W.N. Hardy}
\affiliation{Department of Physics and Astronomy, University of British Columbia, Vancouver BC, V6T 1Z1, Canada}
\author{R.S. Hayano}
\affiliation{Department of Physics, University of Tokyo, Tokyo 113-0033, Japan}
\author{M.E. Hayden}
\affiliation{Department of Physics, Simon Fraser University, Burnaby BC, V5A 1S6, Canada}
\author{A. Humphries}
\affiliation{Department of Physics, Swansea University, Swansea SA2 8PP, United Kingdom}
\author{R. Hydomako}
\affiliation{Department of Physics and Astronomy, University of Calgary, Calgary AB, T2N 1N4, Canada}
\author{S. Jonsell}
\affiliation{Department of Physics, Swansea University, Swansea SA2 8PP, United Kingdom}
\affiliation{Fysikum, Stockholm University, SE-10609, Stockholm, Sweden}
\author{L. Kurchaninov}
\affiliation{TRIUMF, 4004 Wesbrook Mall, Vancouver BC, V6T 2A3, Canada}
\author{R. Lambo}
\affiliation{Instituto de F\'{i}sica, Universidade Federal do Rio de Janeiro, Rio de Janeiro 21941-972, Brazil}
\author{N. Madsen}
\affiliation{Department of Physics, Swansea University, Swansea SA2 8PP, United Kingdom}
\author{S. Menary}
\affiliation{Department of Physics and Astronomy, York University, Toronto, ON, M3J 1P3, Canada}
\author{P. Nolan}
\affiliation{Department of Physics, University of Liverpool, Liverpool L69 7ZE, United Kingdom}
\author{K. Olchanski}
\affiliation{TRIUMF, 4004 Wesbrook Mall, Vancouver BC, V6T 2A3, Canada}
\author{A. Olin}
\affiliation{TRIUMF, 4004 Wesbrook Mall, Vancouver BC, V6T 2A3, Canada}
\author{A. Povilus}
\affiliation{Department of Physics, University of California at Berkeley, Berkeley, CA 94720-7300, USA}
\author{P. Pusa}
\affiliation{Department of Physics, University of Liverpool, Liverpool L69 7ZE, United Kingdom}
\author{F. Robicheaux}
\affiliation{Department of Physics, Auburn University, Auburn, AL 36849-5311, USA}
\author{E. Sarid}
\affiliation{Department of Physics, NRCN-Nuclear Research Center Negev, Beer Sheva, IL-84190, Israel}
\author{D.M. Silveira}
\affiliation{Department of Physics, University of Tokyo, Tokyo 113-0033, Japan}
\affiliation{Atomic Physics Laboratory, RIKEN, Saitama 351-0198, Japan}
\author{C. So}
\affiliation{Department of Physics, University of California at Berkeley, Berkeley, CA 94720-7300, USA}
\author{J.W. Storey}
\affiliation{TRIUMF, 4004 Wesbrook Mall, Vancouver BC, V6T 2A3, Canada}
\author{R.I. Thompson}
\affiliation{Department of Physics and Astronomy, University of Calgary, Calgary AB, T2N 1N4, Canada}
\author{D.P. van der Werf}
\affiliation{Department of Physics, Swansea University, Swansea SA2 8PP, United Kingdom}
\author{D. Wilding}
\affiliation{Department of Physics, Swansea University, Swansea SA2 8PP, United Kingdom}
\author{J.S. Wurtele}
\affiliation{Department of Physics, University of California at Berkeley, Berkeley, CA 94720-7300, USA}
\author{Y. Yamazaki}
\affiliation{Atomic Physics Laboratory, RIKEN, Saitama 351-0198, Japan}
\collaboration{ALPHA Collaboration}
\noaffiliation

\date{Received \today}

\begin{abstract}
We report the application of evaporative cooling to clouds of trapped antiprotons, resulting in plasmas with measured temperature down to 9~K. We have modeled the evaporation process for charged particles using appropriate rate equations. Good agreement between experiment and theory is observed, permitting prediction of cooling efficiency in future experiments.  The technique opens up new possibilities for cooling of trapped ions and is of particular interest in antiproton physics, where a precise CPT test on trapped antihydrogen is a long-standing goal.
\end{abstract}

\maketitle

Historically, forced evaporative cooling has been successfully applied to trapped samples of neutral particles \cite{Hess:1986qc}, and remains the only route to achieve Bose-Einstein condensation in such systems \cite{Anderson:1995fk}. However, the technique has only found limited applications for trapped ions (at temperatures $\sim100$~eV \cite{Kinugawa:2001sw}) and has never been realized in cold plasmas. Here we report the application of forced evaporative cooling to a dense ($\sim10^6$~cm$^{-3}$) cloud of trapped antiprotons, resulting in temperatures as low as 9~K; two orders of magnitude lower than any previously reported \cite{Gabrielse:1989bd}.

The process of evaporation is driven by elastic collisions that scatter high energy particles out of the confining potential, thus decreasing the temperature of the remaining particles. For charged particles the process benefits from the long range nature of the Coulomb interaction, and compared to neutrals of similar density and temperature, the elastic collision rate is much higher, making cooling of much lower numbers and densities of particles feasible. In addition, intraspecies loss channels from inelastic collisions are non-existent. Strong coupling to the trapping fields makes precise control of the confining potential more critical for charged particles. Also, for plasmas, the self-fields can both reduce the collision rate through screening and change the effective depth of the confining potential.

\begin{figure}
   \centering
   \includegraphics[width=\linewidth]{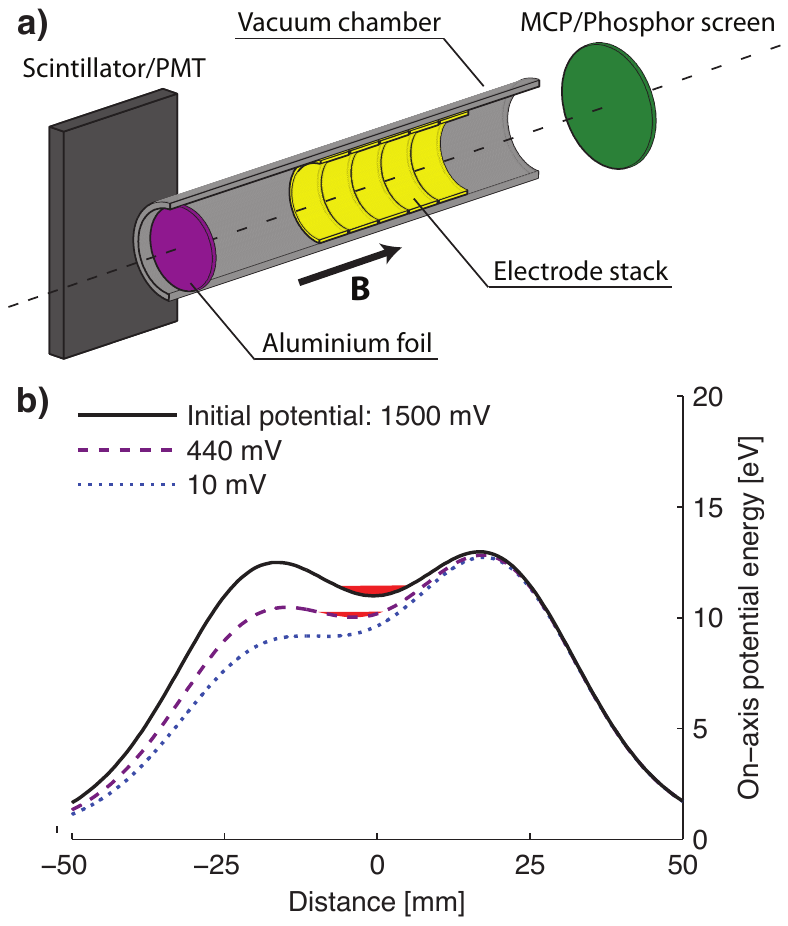} 
      \caption{a) Simplified schematic of the Penning-Malmberg trap used to confine the antiprotons and of the two diagnostic devices used. The direction of the magnetic field is indicated by the arrow. b) Potential wells used to confine the antiprotons during the evaporative cooling ramp. The antiprotons are indicated at the bottom of the potential well (red), and the different wells are labelled by their on-axis depth.}
   \label{fig:ALPHA_apparatus}
\end{figure}

The ALPHA apparatus, which is designed with the intention of creating and trapping antihydrogen \cite{Bertsche:2006xd}, is located at the Antiproton Decelerator (AD) at CERN \cite{Maury:1997yb}. It consists of a Penning-Malmberg trap for charged particles with an octupole-based magnetostatic trap for neutral atoms superimposed on the central region. For the work presented here, the magnetostatic trap was not energized and the evaporative cooling was performed in a homogeneous 1~T solenoidal field.

Figure~\ref{fig:ALPHA_apparatus}a shows a schematic diagram of the apparatus, with only a subset of the 20.05~mm long and 22.275~mm radius, hollow cylindrical electrodes shown. The vacuum wall is cooled using liquid helium, and the measured electrode temperature is about 7~K. The magnetic field, indicated by the arrow, is directed along the axis of cylindrical symmetry and confines the antiprotons radially: due to conservation of angular momentum, antiprotons do not readily escape in directions transverse to the magnetic field lines \cite{ONeil:1980yo}. Parallel to the magnetic field, antiprotons are confined by electric fields generated by the electrodes. 

Also shown are the two diagnostic devices used in the evaporative cooling experiments. To the left, antiprotons can be released towards an aluminium foil, on which they annihilate. The annihilation products are detected by a set of plastic scintillators read out by photomultiplier tubes (PMT) with an efficiency of (25~$\pm$~10)\% per event. The background signal from cosmic rays is measured during each experimental cycle and is approximately 40 Hz. To the right, antiprotons can be released onto a microchannel plate (MCP)/phosphor screen assembly, allowing measurements of the antiproton cloud's spatial density profile, integrated along the axis of cylindrical symmetry \cite{Andresen:2009xz,Andresen:2008ya}.

Before each evaporative cooling experiment a cloud of $45,000$ antiprotons with a radius of $0.6$~mm and a density of $7.5\times10^6$~cm$^{-3}$ is prepared. The antiprotons are produced and slowed to 5.3~MeV in the AD, and as they enter our apparatus through the aluminium foil (see figure \ref{fig:ALPHA_apparatus}a) they are further slowed. Inside the apparatus we capture 70,000 of the antiprotons in a 3~T magnetic field between two high voltage electrodes (not shown) excited to 4~kV \cite{Gabrielse:1986jo}. Typically 65\% of the captured antiprotons spatially overlap a 0.5~mm radius, pre-loaded, electron plasma with 15~million particles. The electrons are self-cooled by cyclotron radiation and in turn cool the antiprotons through collisions \cite{Gabrielse:1989bd}. Antiprotons which do not overlap the electron plasma remain energetic and are lost when the high voltage is lowered.

The combined antiproton and electron plasma is then compressed radially using an azimuthally segmented electrode (not shown) to apply a "rotating-wall" electric field, thus increasing the antiproton density \cite{Andresen:2008ya,Huang:1997hs,Greaves:2000et}. The magnetic field is then reduced to 1~T and the particles moved to a region where a set of low-noise amplifiers is used to drive the confining electrodes.
Pulsed electric fields are employed to selectively remove the electrons, which, being the lighter species, escape more easily. The antiprotons remain in a potential well of depth 1500~mV (see Figure \ref{fig:ALPHA_apparatus}b). The quoted depth is the on-axis value due to the confining electrodes only. Space charge potentials are considered separately. This convention is used throughout the article.

\begin{figure}
   \centering
   \includegraphics[width=\linewidth]{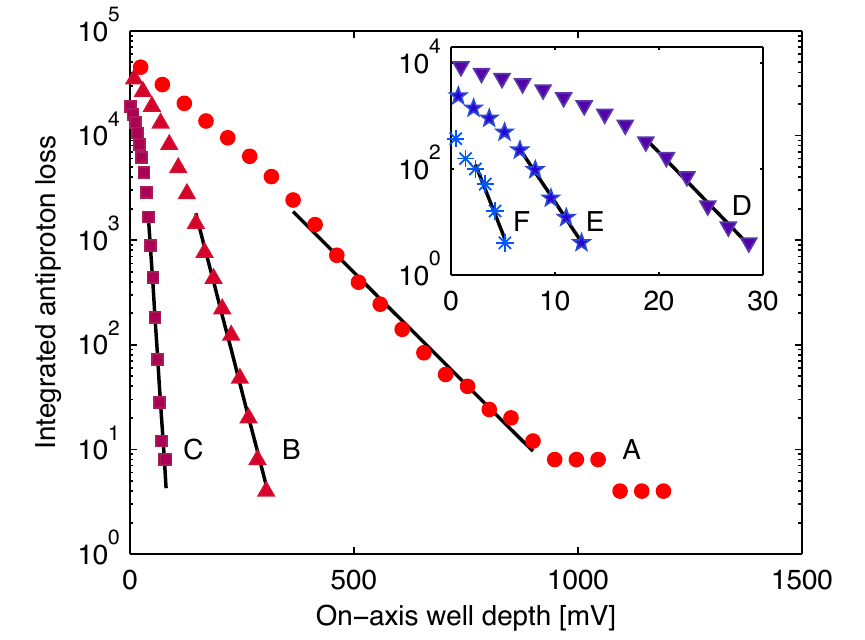} 
      \caption{The number of antiprotons lost from the well as it is lowered is integrated and plotted versus the well depth; note that, in time, the potentials are ramped from right to left in the figure. The measured number is corrected for the 25\% detection efficiency. The curves are labelled in decreasing order of the temperatures extracted from an exponential fit, shown as the solid lines. The temperatures (corrected as described in the text) are: A: 1,040~K B: 325~K C: 57~K D: 23~K E: 19~K F: 9~K. As the antiprotons get colder, fewer can be used to determine their temperature, an effect described in Ref. \cite{Eggleston:1992if}.}
   \label{fig:EVC_EnergyDumps}
\end{figure}

The temperature of the antiprotons can be determined by a destructive measurement in which one side of the confining potential is lowered and the antiprotons released (in a few ms) onto the aluminium foil. Assuming that the particle cloud is in thermal equilibrium, the particles that are initially released originate from the exponential tail of a Boltzmann distribution \cite{Eggleston:1992if}, so that a fit can be used to determine the temperature of the particles. Figure~\ref{fig:EVC_EnergyDumps} shows six examples of measured antiproton energy distributions. 

The raw temperature fits in Figure~\ref{fig:EVC_EnergyDumps} are corrected by a factor determined by particle-in-cell (PIC) simulations of the antiprotons being released from the confining potential. The simulations include the effect of the time dependent vacuum potentials and plasma self-fields, the possibility of evaporation, and energy exchange between the different translational degrees of freedom. The simulations suggest that the temperature determined from the fit is $\sim$16\% higher than the true temperature. Note that the PIC-based correction has been applied to all temperatures reported in this paper. The distribution labelled A in Figure~\ref{fig:EVC_EnergyDumps} yields a corrected temperature of (1040~$\pm$~45)~K before evaporative cooling; the others are examples of evaporatively cooled antiprotons achieved as described below.

To perform evaporative cooling, the depth of the initially 1500~mV deep confining well was reduced by linearly ramping the voltage applied to one of the electrodes to one of six different predetermined values (see examples on Figure~\ref{fig:ALPHA_apparatus}b). Then the antiprotons were allowed to re-equilibrate for 10~s before being ejected to measure their temperature and remaining number. The shallowest well investigated had a depth of (10~$\pm$~4)~mV. Since only one side of the confining potential is lowered, the evaporating  antiprotons are guided by the magnetic  field onto the aluminium foil, where they annihilate. Monitoring the annihilation signal allows us to calculate the number of antiprotons remaining at any time by summing all  antiproton losses and subtracting the measured cosmic background.

\begin{figure}
   \centering
   \includegraphics[width=\linewidth]{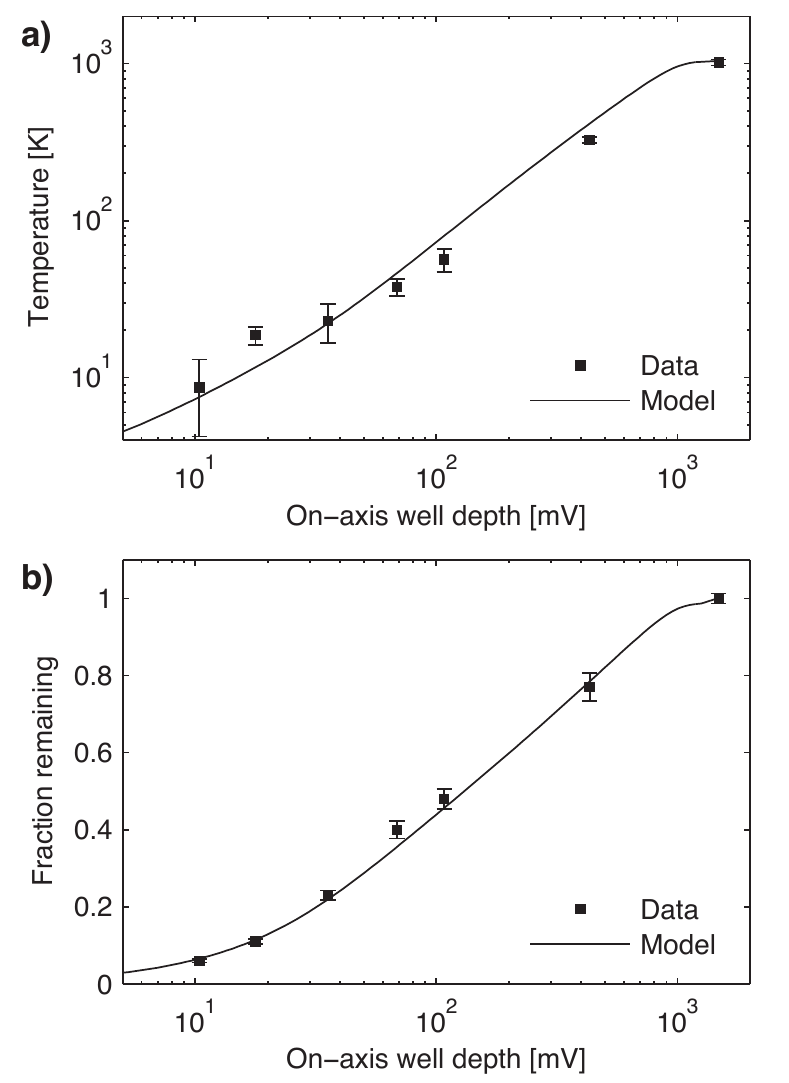} 
      \caption{ a) Temperature vs.\ the on-axis well depth. The error is the combined statistical uncertainty from the temperature fit and an uncertainty associated with the applied potentials (one $\sigma$). The model calculation described in the text is shown as a line. b) The fraction of antiprotons remaining after evaporative cooling vs.\ on-axis well depth. The uncertainty on each point is propagated from the counting error (one $\sigma$). The initial number of antiprotons was approximately 45,000, for an on-axis well depth of (1484~$\pm$~14) mV.}
   \label{fig:EVC_results}
\end{figure}

\begin{figure}
   \centering
   \includegraphics[width=\linewidth]{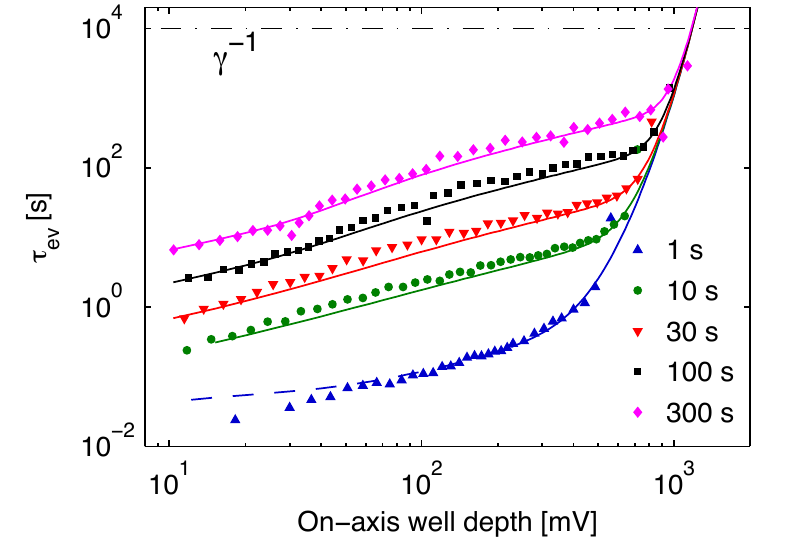} 
      \caption{The evaporation timescale vs.\ applied on-axis well depth for five experiments with different ramp times. The on-axis well depth at the end of the ramp is (10~$\pm$~4)~mV in all cases. $\gamma^{-1}$ is indicated by the dot-dashed line, model calculations are shown as lines and, dashed lines indicate $\eta<4$ in the model calculation.}
      \label{fig:EVC14_results}
\end{figure}

Figure~\ref{fig:EVC_results}a shows the temperature obtained during evaporative cooling as a function of the well depth. We observe an almost linear relationship, and in the case of the most shallow well, we estimate the temperature to be (9~$\pm$~4)~K. The fraction of antiprotons remaining at the various well depths is shown on Figure~\ref{fig:EVC_results}b, where it is found that (6~$\pm$~1)\% of the initial 45,000~antiprotons remain in the shallowest well.

We investigated various times (300~s, 100~s, 30~s, 10~s, 1~s) for ramping down the confining potential from 1500~mV to 10~mV. The final temperature and fraction remaining were essentially independent of this time except for the 1~s case, for which only 0.1\% of the particles survived.

A second set of measurements was carried out to determine the transverse antiproton density profile as a function of well depth. For these studies the antiprotons were released onto the MCP/phosphor screen assembly (see Figure~\ref{fig:ALPHA_apparatus}a), and the measured line-integrated density profile was used to solve the Poisson-Boltzmann equations to obtain the full three dimensional density distribution and electric potential \cite{Spencer:1993kc}.

A striking feature of the antiproton images was the radial expansion of the cloud with decreasing well depth, from an initial radius $r_0$ of 0.6~mm to approximately 3~mm for the shallowest well. If one assumes that all evaporating antiprotons are lost on the axis, where the confining electric field is weakest, no angular momentum is carried away in the loss process. Conservation of total canonical angular momentum \cite{ONeil:1980yo} would then predict that the radial expansion of the density profile will follow the expression $N_0/N=\langle r^2\rangle/\langle r_0^2\rangle$ when angular momentum is redistributed among fewer particles into a new equilibrium. Here $N_0$ is the initial number of antiprotons and $N$ and $r$ are, respectively the number and radius after evaporative cooling. We find that this simple model describes the data reasonably well.

To predict the effect of evaporative cooling in our trap we modelled the process by solving the rate equations describing the time evolution of the temperature $T$, and the number of particles trapped $N$ \cite{Ketterle:1996zv}:
\begin{align}
	\frac{dN}{dt}&=-\frac{N}{\tau_{ev}}-\gamma N, \label{eq:DiffEqN}\\  
	\frac{dT}{dt}&=-\alpha\frac{T}{\tau_{ev}}+P, \label{eq:DiffEqT}
\end{align}
where $\tau_{ev}$ is the evaporation timescale, and $\alpha$ the excess energy removed per evaporating particle. A loss term $\gamma$ ($1\times10^{-4}$ s$^{-1}$ per antiproton) was added to allow for the measured rate of antiproton annihilation on the residual gas in the trap, and a heating term $P$ was also included to prevent the predicted temperatures from falling below the measured limits. The value of $P$ is determined by calculating Joule heating from the release of potential energy during the observed radial expansion. Assuming that the expansion is only due to particle loss and the conservation of angular momentum, as described above, we find $P$ to be of order $(-dN/dt)\times$5~mK. 

Following the notation in Ref.\ \cite{Ketterle:1996zv} we calculate $\alpha$ as:
\begin{equation}
	\alpha=\frac{\eta+\kappa}{\delta+3/2}-1,
\end{equation}
where $\eta$ is the potential barrier height, $\kappa$ the excess kinetic energy of an evaporating particle and, $\delta + 3/2$ the average potential and kinetic energy of the confined particles. The variables $\eta$, $\kappa$ and $\delta$ are the respective energies divided by $k_BT$, where $k_B$ is Boltzmann's constant. Ignoring the plasma self-potential we let $\delta=1/2$ and, following \cite{Ketterle:1996zv} set $\kappa=1$.

Using the principle of detailed balance, $\tau_{ev}$ can be related to the relaxation time $\tau_{col}$ for antiproton-antiproton collisions with:
\begin{equation}\label{eq:tau_ev}
	\frac{\tau_{ev}}{\tau_{col}}=\frac{\sqrt{2}}{3}\eta\, e^{\eta},
\end{equation}
being the appropriate expression for one dimensional evaporation \cite{Fussmann:1999lo}. Note that this expression is an approximation only valid for $\eta>4$ \cite{Fussmann:1999lo} and, that  the ratio is a factor of $\eta$ higher than the corresponding expression for neutrals \cite{Ketterle:1996zv}. For $\tau_{col}$ we use the expression calculated in Ref.\ \cite{Glinsky:1992zh}. 

From the antiproton density profiles, we estimate the central self-potential to be about 1.5~$\mu$V per antiproton. In the initial 1500~mV well, this can be ignored, but with e.g. 20,000 antiprotons left in a 108~mV well a total self-potential of 30~mV changes our estimate of $\eta$ from about 19 to 14 by reducing the energy required to evaporate from the well. The corresponding reduction in $\tau_{ev}$ is two orders of magnitude. Thus, the effect was included in the model, decreasing the predicted value of $N$ by  almost an order of magnitude in some cases.

Figure~\ref{fig:EVC14_results} shows a comparison between the measured and the predicted evaporation timescales, calculated as $\tau_{ev}=-(d\log N/dt\ +\ \gamma)^{-1}$, for five different ramp times to the most shallow well. The good agreement between the measurement and the model calculations shows that we can predict $\tau_{ev}$ over a wide range of well depth and ramp times, giving good estimates of $N$ in equation (\ref{eq:DiffEqN}). The magnitude of $\gamma^{-1}$ is indicated on the figure, showing that antiproton loss to annihilations on residual gas is a small effect.

In the calculations, shown as lines on Figure~\ref{fig:EVC_results}, we have modelled the temperature and remaining number as a function of the well depth. Of most interest is panel a), where reasonable agreement between the measured and the predicted temperatures validates the choice of $\alpha$ and $P$. With the measurements available we cannot exclude, but only limit, other contributions to $P$ to be of similar magnitude as that from expansion-driven Joule heating. One such contribution could be electronic noise from the electrode driving circuits coupling to the antiprotons. The model also explains the excessive particle loss observed in the 1~s ramp. Here $\tau_{ev}$ becomes too short for rethermalization and $\eta$ is forced below one.

Evaporative cooling represents a strong candidate to replace electron cooling of antiprotons as the final cooling step when preparing pure low temperature samples of antiprotons for production of cold antihydrogen. Such samples can be used either as a target to create antihydrogen through charge exchange with positronium atoms \cite{Charlton:1990fx,Storry:2004sh} or having no intrinsic cooling mechanism they can be precisely manipulated to have a well defined energy relative to an adjacent positron plasma into which they are later injected. The technique eliminates electron related difficulties, such as heating by plasma instabilities \cite{Peurrung:1993ko} and centrifugal separation \cite{{ONeil:1981ko}} in mixed plasmas, absorption of thermal radiation and high frequency noise from the electrode driving circuit, and pulsed-field heating during removal. In principle, temperatures lower than the temperature of the surrounding apparatus are obtainable. Perhaps even sub-Kelvin plasmas, foreseen to be a prerequisite for proposed measurements of gravitational forces on antimatter \cite{Drobychev:2007bx}, can be achieved.  

Despite reducing the total antiproton number, we have increased the absolute number of antiprotons with an energy below our 0.5~K neutral atom trap depth by about two orders of magnitude compared to the initial distribution; from less than one to more than 10 in each experiment. This greatly improves the probability of producing trappable antihydrogen in many proposed antihydrogen production schemes \cite{Amoretti:2002cr,Gabrielse:2002fh,Charlton:1990fx,Storry:2004sh}. Preliminary studies indicate that repeating the experiment with our octupole-based atom trap energized does not change the outcome. Electronic noise in the electrode driving circuit and the ability to precisely set the confining potential will most likely limit the lowest temperature achievable in any given apparatus.

The evaporative cooling of a cloud of antiprotons down to temperatures below 10~K has been demonstrated. In general, the technique could be used to create cold, non-neutral plasmas of particle species that cannot be laser cooled.

This work was supported by CNPq, FINEP (Brazil), ISF (Israel), MEXT (Japan), FNU (Denmark), VR (Sweden), NSERC, NRC/TRIUMF, AIF (Canada), DOE, NSF (USA), EPSRC and the Leverhulme Trust (UK).

\end{document}